\documentclass[referee]{raa}
\usepackage{amssymb,amsmath,graphicx,times,natbib,subfig}
\usepackage[a4paper=true,dvipdfm=true,pagebackref=false]{hyperref}
\hypersetup{pdftitle = The title of my PDF, pdfauthor = My name, pdfsubject= The subject, pdfkeywords = keyword1 keyword2 keyword3}
\hypersetup{colorlinks = true, linkcolor = green, anchorcolor = red, citecolor = blue, filecolor = red, pagecolor = red, urlcolor = red}

\begin{document}

   \title{ A Low-latency Pipeline for GRB Light Curve and Spectrum using {\it Fermi}/GBM Near Real-time Data }

   \volnopage{ {\bf 201x} Vol.\ {\bf XXX} No. {\bf XXX}, 000--000}
   \setcounter{page}{1}
   \author{ Yi Zhao\inst{1,2}, Bin-Bin Zhang\inst{3,4}, Shao-Lin Xiong\inst{2}, Xi Long\inst{2,5}, Qiang Zhang\inst{2}, Li-Ming Song\inst{2}, Jian-Chao Sun\inst{2}, Yuan-Hao Wang\inst{2,5}, Han-Cheng Li\inst{2,5}, Qing-Cui Bu\inst{2}, Min-Zi Feng\inst{2,5}, Zheng-Heng Li\inst{2,5}, Xing Wen\inst{2,5}, Bo-Bing Wu\inst{2}, Lai-Yu Zhang\inst{2}, Yong-Jie Zhang\inst{2}, Shuang-Nan Zhang\inst{2} and Jian-Xiong Shao\inst{1} }

   \institute{ School of Nuclear Science and Technology, Lanzhou University, Lanzhou 730000, China; {\it yizhao@ihep.ac.cn}\\
              \and
               Key Laboratory of Particle Astrophysics, Institute of High Energy Physics, Chinese Academy of Sciences, Beijing 100049, China; {\it xiongsl@ihep.ac.cn}\\
              \and
               School of Astronomy and Space Science, Nanjing University, Nanjing 210093, China\\
              \and
               Instituto de Astrof\'{i}sica de Andaluc\'{i}a (IAA-CSIC), P.O. Box 03004, E-18080 Granada, Spain\\
              \and
               University of Chinese Academy of Sciences, Beijing 100049, China\\
        \vs \no
        {\small Received 2017 December 14; accepted 2018 February 11 }
            }

\abstract{
    Rapid response and short time latency are very important for Time Domain Astronomy, such as the observations of Gamma-ray Bursts (GRBs) and electromagnetic (EM) counterparts of gravitational waves (GWs). Based on the near real-time {\it Fermi}/GBM data, we developed a low-latency pipeline to automatically calculate the temporal and spectral properties of GRBs. With this pipeline, some important parameters can be obtained, such as $T_{90}$ and fluence, within $\sim20$~minutes after the GRB trigger. For $\sim90\%$ GRBs, $T_{90}$ and fluence are consistent with the GBM catalog results within 2~$\sigma$ errors. This pipeline has been used by the Gamma-ray Bursts Polarimeter (POLAR) and the {\it Insight} Hard X-ray Modulation Telescope ({\it Insight}-HXMT) to follow up the bursts of interest. For GRB 170817A, the first EM counterpart of GW events detected by {\it Fermi}/GBM and INTEGRAL/SPI-ACS, the pipeline gave $T_{90}$ and spectral information in 21 minutes after the GBM trigger, providing important information for POLAR and {\it Insight}-HXMT observations.
}

    \authorrunning{ Yi Zhao et al. }                  
    \titlerunning{  A Low-latency Pipeline for GRBs } 
    \maketitle

\section{INTRODUCTION}

    Gamma-Ray Bursts (GRBs) are the catastrophic processes happening during either the violent stellar explosions or the coalescence of binary compact stars in the Universe. Since the discovery of GRBs by the {\it Vela} satellites was published in 1973 (Klebesadel et al.~\citealp{kelbe1973}), GRBs have become a hot topic for astrophysical researches. Thereafter many dedicated gamma-ray space telescopes, such as {\it CGRO}, {\it BeppoSAX}, {\it HETE-2}, {\it Swift} and {\it Fermi}, were launched to explore the properties of GRBs (Boella et al.~\citealp{boell1997}; Gehrels et al.~\citealp{gehre2004}; Lin~\citealp{linyq2009}; Meegan et al.~\citealp{meega2009}). The physics of GRBs, however, is still an open question (Zhang~\citealp{zhang2011}; Kumar \& Bing~\citealp{kumar2015}). Among them, the prompt emission mechanism of GRBs is highly uncertain. Apart from the light curve and spectral analysis, the polarization of $\gamma$-ray photons is of great importance for the constraint of prompt emission mechanism models. However, a precise polarization measurement is still lacking. The Gamma-ray Bursts Polarimeter (POLAR), launched on-board the Chinese Space Laboratory "Tiangong-2" (TG2) on the 15th of September 2016, is a space-borne Compton polarimeter aiming to measure the polarization of GRB prompt emission in the 50--500~keV energy range with high accuracy (Produit et al.~\citealp{produ2005}, ~\citealp{produ2018}; Suarez-Garcia et al.~\citealp{suare2010}; Orsi et al.~\citealp{orsis2011}; Kole et al.~\citealp{kolem2017}).

    Since POLAR is the best GRB polarimeter ever flown, the most interesting GRBs for POLAR are those with high polarization measurement accuracy, i.e. low Minimum Detectable Polarization (MDP) (Davide~\citealp{david2006}; Toma et al.~\citealp{tomak2009}; Xiong et al.~\citealp{xiong2009}). For a given GRB, MDP is determined by the location, $T_{90}$\footnote{$T_{90}$ is defined as the time during which cumulative counts increase from 5\% to 95\% above background (Kouveliotou et al.~\citealp{kouve1993}).} and fluence of the GRB. However, due to the telemetry resource limitation of POLAR, no real-time data is available and the delay would be typically $\sim10$~hours. In order to enhance the science result productions of POLAR, a dedicated real-time alert system has been set up to monitor GRBs reported by other telescopes (e.g. {\it Swift}, {\it Fermi}/GBM) to predict whether they are visible by POLAR based on the orbit of TG2 and evaluate the MDP of those GRBs. Finally, we circulate the interesting GRBs to the GRB community to encourage follow-up observations.

    In order to facilitate follow-up observations, it is important to estimate the MDP with low latency. Thus, we need to get the GRB location, duration and fluence information at the earliest time after the GRB trigger. Meanwhile, the GRB properties mentionded above and other properties like whether a GRB is short or long, and what the spectral hardness is, are also helpful for the detection of the electromagnetic (EM) counterparts of gravitational waves (GWs) (Berger~\citealp{berge2014}). Therefore, the low-latency calculation of $T_{90}$ and fluence for GRBs is highly required.

    So far, the Gamma-ray Burst Monitor (GBM) onboard the {\it Fermi} Gamma-ray Space Telescope is one of the most sensitive instruments for GRB detection. In the past $\sim9$~years, {\it Fermi}/GBM has detected $\sim2200$ GRBs, which means $\sim240$ GRBs per year (Paciesas et al.~\citealp{pacie2012}; Kienlin et al.~\citealp{kienl2014}; Bhat et al.~\citealp{bhatp2016}). In addition, {\it Fermi}/GBM sends a series of GCN Notice\footnote{https://gcn.gsfc.nasa.gov/fermi\_grbs.html} (including FLT, GND and FINAL Notices) within several minutes and releases the near real-time data within $\sim15$~minutes after the GBM trigger. Although these GCN Notices can report the trigger time, the GRB location, and the link to the near real-time data with counts rate (i.e. light curve) in 8 energy channels, etc., no direct information about $T_{90}$ and spectral parameters are provided.

    To satisfy the requirement of POLAR, we developed a low-latency pipeline, based on the near real-time data of {\it Fermi}/GBM, to automaticly calculate the temporal and spectral properties of GRBs, including $T_{90}$ and fluence.

    This paper is structured as follows. In Section 2, the pipeline including the data selection and reduction is briefly described. Then the performance is validated by comparing the $T_{90}$ and spectral results analyzed by the pipeline with those of GBM GRB catalog for 700 GRBs in Section 3. Finally, in Section 4, we summarize our results and make a brief discussion. Unless otherwise stated, all errors adopted in this paper are given at the 1~$\sigma$ confidence level.

\section{DESCRIPTION OF THE PIPELINE}
\subsection{Data Selection}

    The original GBM observational data are available online on the GBM data server\footnote{https://heasarc.gsfc.nasa.gov/FTP/fermi/data/gbm/bursts/}. Three types of information are required by the pipeline: science data, GRB location and GBM Detector Response Matrices (DRMs). For each burst, GBM provides three types of science data: CTIME, CSPEC and Time-Tagged Events (TTE) data. The CTIME data has fine time resolution but coarse spectral resolution. In contrast, the CSPEC data has coarse time resolution but fine spectral resolution. The TTE data consists of individual photon events with fine time resolution (2 microseconds) and fine spectral resolution (128 energy channels). In practice, GBM usually use the CTIME or TTE data for $T_{90}$ calculation and use the CSPEC or TTE data for spectral analysis (Meegan et al.~\citealp{meega2009}).

\begin{figure}[!ht]
    \centering
    \includegraphics[width=\textwidth, angle=0]{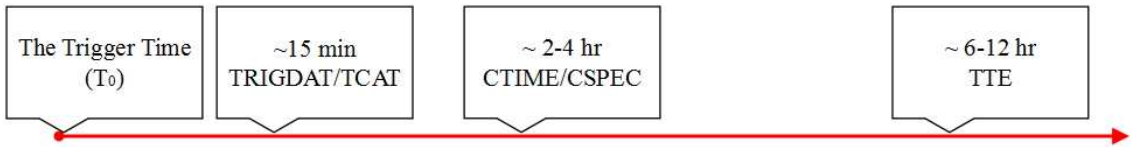}
    \caption{ Timeline of the {\it Fermi}/GBM data products. Our pipeline employs the TRIGDAT data as the science data and the TCAT data as the location files for the sake of timeliness. }
    \label{raa20170280r1_fig1}
    \end{figure}

    Once GBM is triggered by a burst, the TRIGDAT data will be immediately downlinked through the Tracking and Data Relay Satellite System (TDRSS) in near-real time. Then it will be quickly available on the GBM data server in $\sim15$~minutes after the trigger. The TRIGDAT data includes the counts rate of all 14 GBM detectors with coarse spectral resolution of 8 energy channels, spanning from $T_0-150$~s to $T_0+450$~s. For the TRIGDAT data, the bin size of the counts rate is not constant. It is typically 1.024 s, and sometimes it is 0.064~s or 0.256~s for only 1--2~bins in the time region from $T_0$ to $T_0+60$~s, while for other regions it is 8.192~s (Meegan et al.~\citealp{meega2009}). Although the TRIGDAT data is not as precise as the CTIME, CSPEC and TTE data, it can be available within about 15 minutes after the GRB trigger, which is much faster than other types of data, as shown in Figure~\ref{raa20170280r1_fig1}. Therefore, the TRIGDAT data is chosen as the science data input of the pipeline.

    GBM circulates the GRB location information (including flight, ground and final locations) via the GCN Notice. The final location information is recorded in the TCAT data. The TCAT and TRIGDAT data usually appear on the data server at the same time. Therefore, our pipeline employs the TCAT data as location files.

    The DRMs are mandatory for spectral analysis. However, the GBM-released DRM files are not available before the CTIME and CSPEC data arrive. As part of the science analysis tool provided by the GBM, the response matrix generator\footnote{https://fermi.gsfc.nasa.gov/ssc/data/analysis/rmfit/gbmrsp-2.0.10.tar.bz2} is designated to generate the DRMs according to the trigger time and the attitude information of the spacecraft. Thus the response matrix generator is utilized by our pipeline to generate the DRMs.

\subsection{Data Reduction}

\begin{figure}[!ht]
    \centering
    \includegraphics[width=\textwidth, angle=0]{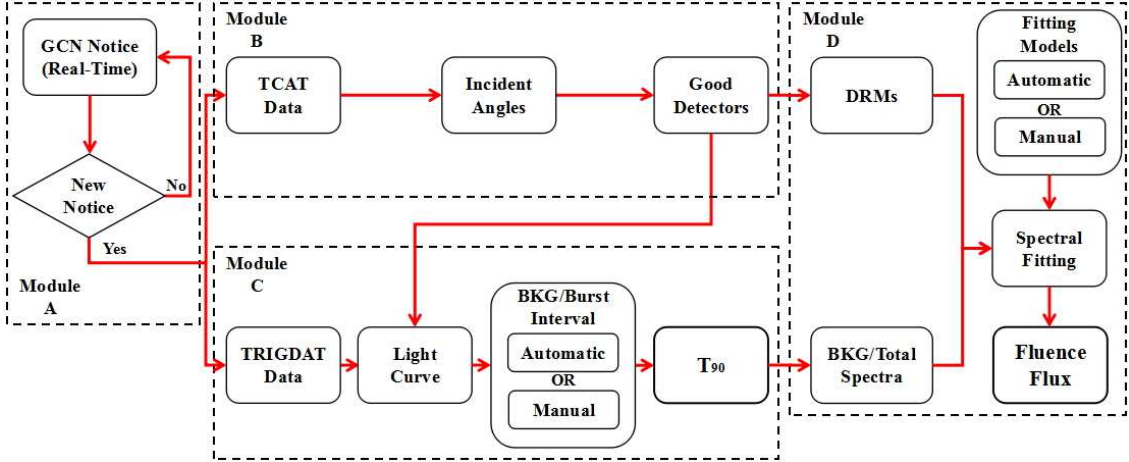}
    \caption{ The design of our low-latency light curve and spectral analysis pipeline. }
    \label{raa20170280r1_fig2}
    \end{figure}

    As shown in Figure~\ref{raa20170280r1_fig2}, our low-latency light curve and spectral analysis pipeline mainly consists of 4 modules: A, B, C and D. The function of each module is described below. Detailed analysis methods are presented in the following subsections.

    Module A: monitor the GCN Notice. The pipeline monitors the GCN Notices released by GBM for new-detected GRBs in real time. Once a new GBM GCN Notice arrives, the corresponding TRIGDAT and TCAT data will be automatically downloaded.

    Module B: select the good detectors. The incident angles of 12 Sodium Iodide (NaI) and 2 Bismuth Germinate (BGO) detectors are computed according to the GRB location and the spacecraft attitude recorded in the TCAT data. Then the two NaI detectors with the minimum incident angles which are less than 60 deg and the BGO detector with the optimum incident angle (the so-called `good detectors') are selected for the $T_{90}$ and spectral analysis in the pipeline. Meanwhile, the DRMs can be calculated by the response matrix generator.

    Module C: calculate $T_{90}$. We note that the time duplicate data exist in the TRIGDAT files, which have been removed by the pipeline before using. Once the light curve is generated, the background and burst intervals will be automatically selected through the iteration method (see Section 2.3). Fitting and subtracting the background will be followed, and then $T_{90}$ will be calculated. Our pipeline also provides an option for users to choose the background and source intervals manually.

    Module D: spectral analysis. The pipeline automatically analyze the total spectra and background spectra according to the obtained $T_{90}$. Using the spectra and the DRM files, this pipeline will automatically fit the observational data and calculate the fluence. Complementary to the automatic fitting, manually selection of the fitting models is also provided.

    Generally, the GBM near real-time data (TRIGDAT, TCAT) can be available within $\sim15$~minutes. For most GRBs, $T_{90}$ will be automatically calculated within 30~s and the spectral analysis results can be obtained within $\sim5$~minutes. Therefore, our pipeline can give the results within $\sim20$~minutes after trigger.

    The pipeline will push the obtained results of $T_{90}$ and spectral parameters to the internal web server of POLAR. These results can also be obtained by the POLAR Burst Advocates (BA) through manually selecting the background intervals, the burst intervals and the fitting models, as shown in Figure~\ref{raa20170280r1_fig2}. In this case, another several minutes are needed considering the BA response.

\subsection{$T_{90}$ calculation}

\begin{figure}[!ht]
    \centering
    \includegraphics[width=9cm, angle=0]{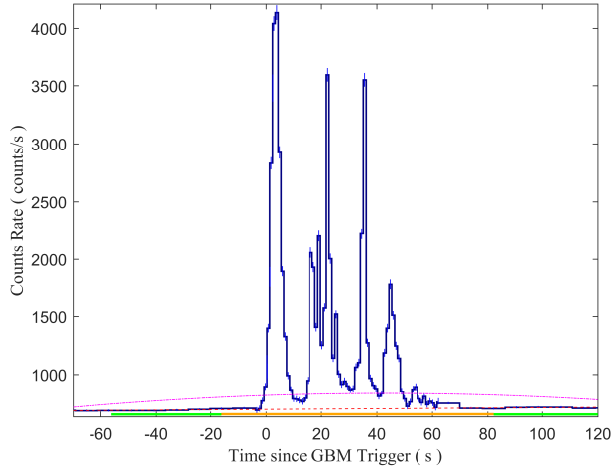}
    \caption{ Illustration of automatic $T_{90}$ calculation, taking bn090626189 as an example. The magenta dot-dashed line is the initial background fitting curve and the red dashed line is the background fitting curve. The orange horizontal line is the burst interval used to calculate $T_{90}$, and the two green horizontal lines on both sides of the orange horizontal line are the final background intervals. }
    \label{raa20170280r1_fig3}
    \end{figure}

    This pipeline calculates $T_{90}$ in the 50--300~keV energy range using the two NaI detectors. The main steps of automatic $T_{90}$ calculation using the the iteration method are as follows. First of all, our pipeline employs a quadratic polynomial function to fit the whole light curve, and then obtains the initial background fitting curve, as shown by the magenta dot-dashed line in Figure~\ref{raa20170280r1_fig3}. The data points, which are less than 3~$\sigma$ from the initial fitting curve, will be used as the updated background intervals.

    The pipeline uses a quadratic polynomial function to fit the updated background intervals, and then obtain the updated background fitting curve. After that, the pipeline will seek for the data points of the light curve which deviate the updated background fitting curve by more than 4~$\sigma$. Based on the time interval of these data points, our pipeline will extend some time margin on both sides to include a part of background as the burst interval, as shown by the orange horizontal line in Figure~\ref{raa20170280r1_fig3}. Here we choose 18~s after a large amount of test in the pipeline. The final background intervals are two 40 seconds intervals on both sides of the burst interval as shown by the two green horizontal lines in Figure~\ref{raa20170280r1_fig3}. We found that the burst interval for most GRBs can be selected reasonably well by iterating once after a large quantity of test (see Section 3.1). Thus we only iterate once for the sake of determining the final background intervals.

    The background eventually adopted by our pipeline, as shown by the red dashed line in Figure~\ref{raa20170280r1_fig3}, is given by a linear fitting to the final background intervals. The $T_{90}$ is calculated by accumulating the counts through the burst interval 5\% to 95\% above the background (Kouveliotou et al.~\citealp{kouve1993}), which is different from $T_{90}$ calculated by fluence accumulation in GBM (Goldstein et al.~\citealp{golds2012}). However, the comparison of $T_{90}$~s suggests that $T_{90}$~s calculated by the pipeline are basically consistent with those given by GBM (see Section 3.1). The error of $T_{90}$ is computed by performing Monte-Carlo simulations considering statistical fluctuation of all data points during the burst interval. Only the Poisson error of data points caused by the statistical fluctuation is considered.

    Following this method, we analyzed bn090626189 as an example. Our automaticly calculated $T_{90}$ result of bn090626189 is consistent with the GBM result at 1~$\sigma$ level, as tabulated in Table~\ref{Tab:T1}.

    \begin{table}[!ht]
    \begin{center}
    \caption[]{ The calculation result of bn090626189. }\label{Tab:T1}
    \begin{tabular}{cccc}
        \hline\noalign{\smallskip}
                                  &  $T_{90}$ (s)        &  $T_{90}$ Start (s)  &  $T_{90}$ End (s)  \\
        \hline\noalign{\smallskip}
            Our Automatic Result  &  $ 46.46 \pm 1.54 $  &    $T_0+1.41$        &  $T_0+47.87$  \\
            GBM Result            &  $ 48.90 \pm 2.83 $  &    $T_0+1.54$        &  $T_0+50.43$  \\
        \noalign{\smallskip}\hline
    \end{tabular}
    \end{center}
    \end{table}

    We emphasize that the difference between manual and automatic calculations of $T_{90}$ is obvious. The former selects the background and burst intervals manually, while the later automatically. The comparison of our automatically obtained $T_{90}$ with those given by GBM will be discussed in Section 3.

\subsection{Spectral Analysis}

    Using the background and $T_{90}$ intervals calculated by the pipeline, the spectral analysis can be conducted. We performed the energy spectral fitting using {\it McSpecFit} (Zhang et al.~\citealp{zhang2016a}), which is a software package that combines a Bayesian Monte-Carlo (MC) engine {\it McFit}, the general forward-folding algorithms and the likelihood calculations. Particularly, the Bayesian Monte-Carlo engine ({\it McFit}) employs a Bayesian MC fitting algorithm to precisely fit the spectra (Zhang et al.~\citealp{zhang2016b}). Previous researches show that the cutoff power law (CPL) model is preferred by most GRBs (e.g., Goldstein et al.~\citealp{golds2012}; Gruber et al.~\citealp{grube2014}; Yu et al.~\citealp{yuhoi2016}), therefore we choose the CPL model in our pipeline to fit the spectra by default,
    \begin{equation}
    M_{\rm CPL}(E,P) = A (\frac{E}{E_{\rm piv}})^{\alpha} exp[-\frac{(\alpha+2)E}{E_{\rm peak}}]
    \end{equation}
where A is the amplitude, $\alpha$ is the low-energy spectral index, $E_{\rm peak}$ is the peak energy, and $E_{\rm piv}$ is fixed to 100~keV. If the fitting fails, a power law model or a Band function (Band et al.~\citealp{bandd1993}) will be used. The priors of all fitting free parameters are set to uniform distribution, with only exception of the amplitude which is set to log distribution. With this algorithm, the best-fit spectral parameters and their uncertainties could be calculated by the converged MC chains. Thus the general forward-folding algorithms are used to compare a spectral model to data.

\begin{figure}[!ht]
    \centering
    \subfloat[]{%
    \includegraphics[width=6cm]{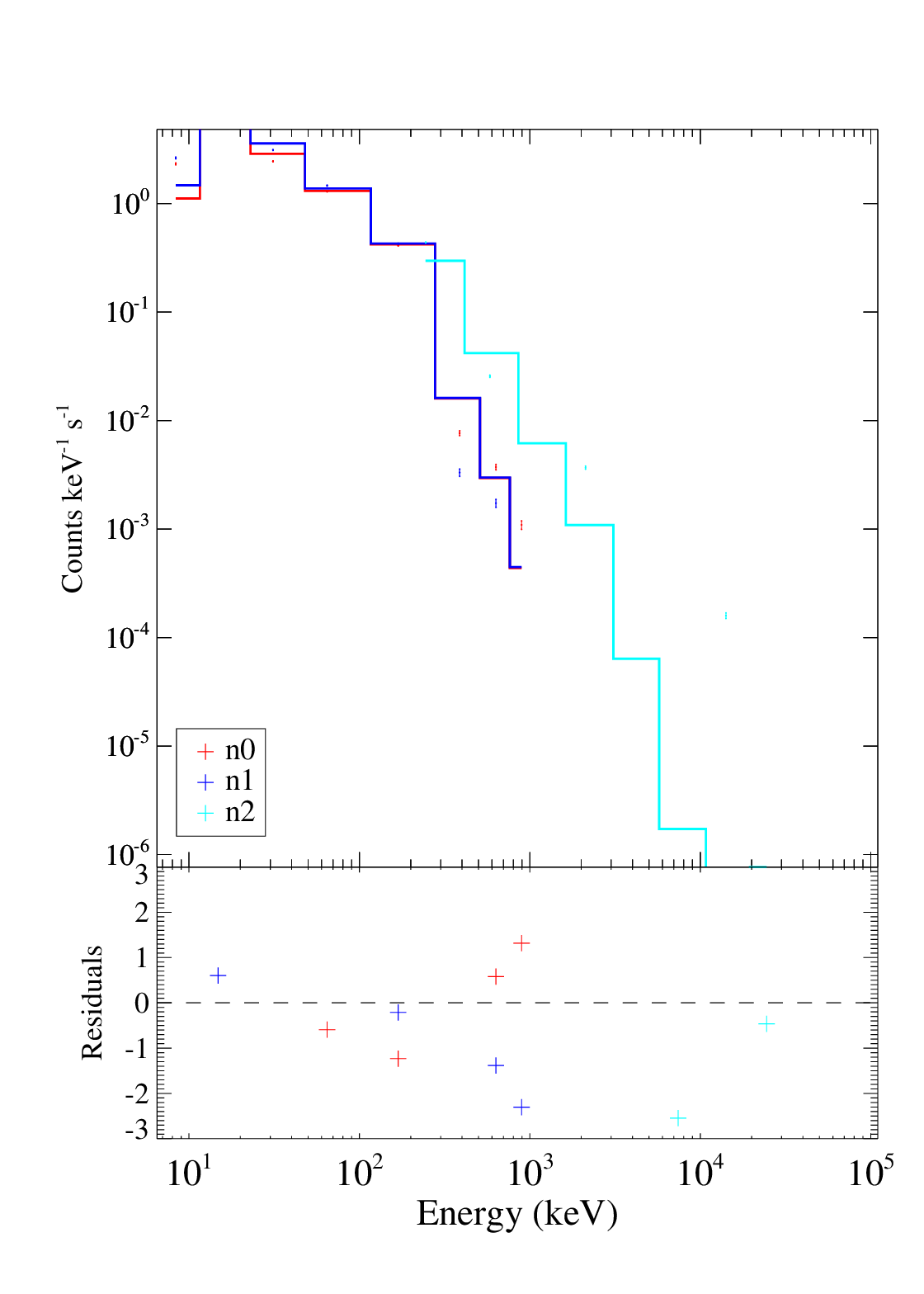}}\hfill
    \subfloat[]{%
    \includegraphics[width=8cm]{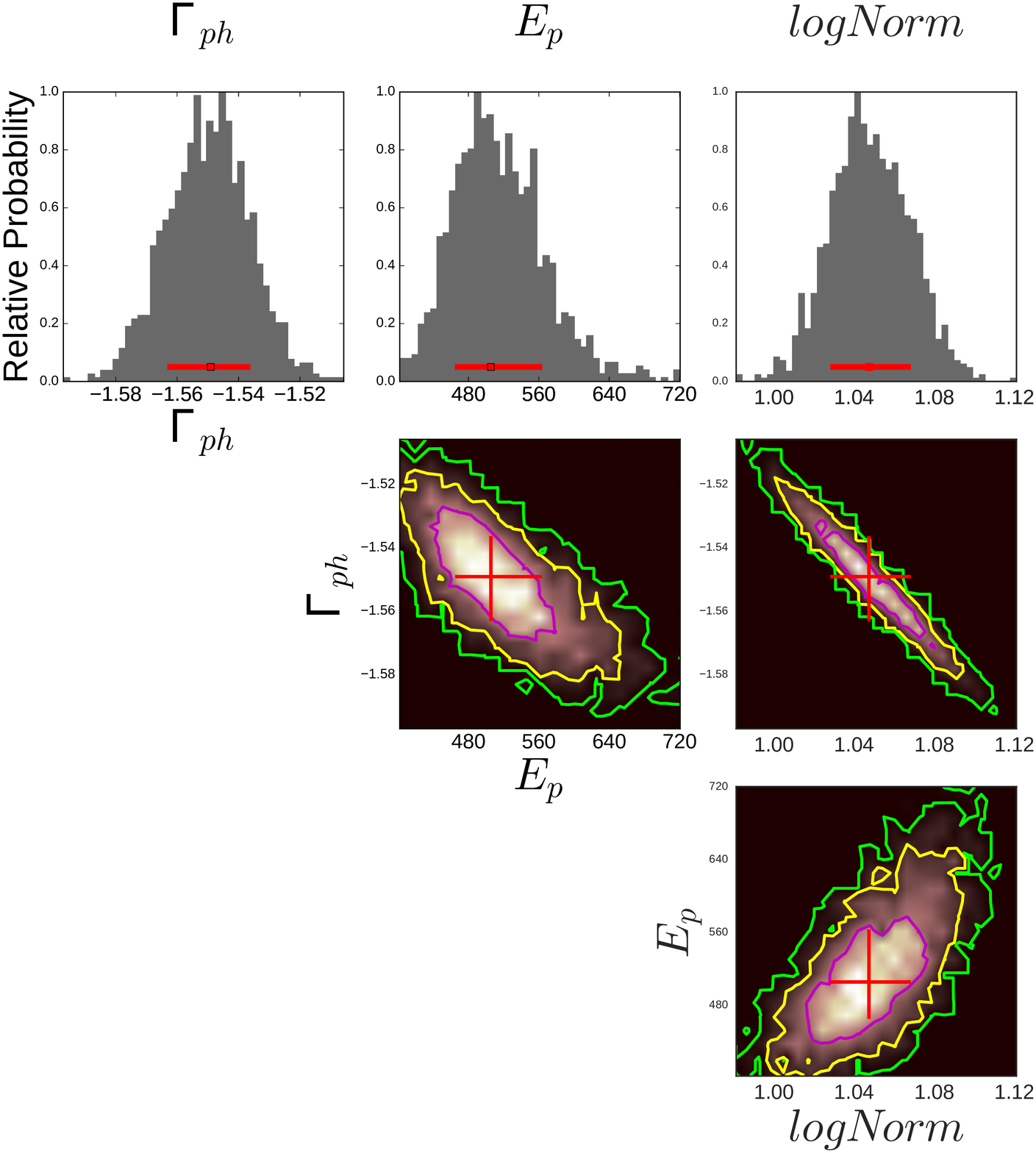}}\hfill
    \caption{ Illustration of spectral fitting of bn090626189 with the CPL model. Panel (a): joint spectral fitting with the good detectors. Panel (b): the posterior corner maps given by {\it McSpecFit}. $\Gamma_{ph}$, $E_{p}$ and $logNorm$ are the low-energy spectral index, the peak energy and the logarithmic amplitude of the CPL model, respectively. }
    \label{raa20170280r1_fig4}
    \end{figure}

    Then {\it McSpecFit} convolves the model with the DRMs to compare spectral models with the net counts spectra,
    \begin{equation}
    C_{M}(I,P) = \int M(E,P) D(I,E) dE
    \end{equation}
where $C_{M}(I,P)$ is the model-predicted count spectrum, $M(E,P)$ is the model spectrum and $D(I,E)$ represents the DRM. A forward-folding algorithm is used to deal with the GBM response $D(I,P)$ and read in the model spectra $M(E,P)$, then the count spectra will be fit. $C_{M}(I,P)$ can be directly compared with the observed count spectra $C_{O}(I)$, as shown in Figure~\ref{raa20170280r1_fig4} (a). Afterwards, {\it McSpecFit} calculates the likelihood for those $C_{M}(I,P)$ and $C_{O}(I)$ pairs. In the pipeline, the maximum likelihood-based statistics for Poisson data (Cash.~\citealp{cashw1979}) with Gaussian background (i.e. PGSTAT\footnote{See also https://heasarc.gsfc.nasa.gov/xanadu/xspec/manual/XSappendixStatistics.html .}) is used to estimate fitting parameters (see Figure~\ref{raa20170280r1_fig4} (b)).

    Finally the fluence is calculated from the integration of the spectral model in 10--1000~keV. The error of the fluence is calculated with Monte-Carlo simulations.

    Differently, the manual spectral analysis selects the background and burst intervals by the BA, and the fitting models can be recognized manually. The analyzed results of automatic and manual calculations of fluence will be discussed in Section 3.

\section{PERFORMANCE OF THE PIPELINE}

    To examine the performance of this pipeline, we choose 700 {\it Fermi} GRBs before bn121211574 from the GBM catalog to test the reliability of the results. The TRIGDAT data of these GRBs meets the following conditions: (1) no less than three data points in the background intervals; (2) no less than one data point in the burst interval. The background and burst intervals are from the GBM catalog. In general, the pipeline takes $\sim20$~seconds to get $T_{90}$ and $\sim5$~minutes to obtain spectral fitting parameters and fluence. We compared our calculated results of $T_{90}$ and fluence with the GBM catalog results (Goldstein et al.~\citealp{golds2012}; Gruber et al.~\citealp{grube2014}; Bhat et al.~\citealp{bhatp2016}).

    The comparisons are shown in Figure~\ref{raa20170280r1_fig5} and Figure~\ref{raa20170280r1_fig8}. If our results are essentially in agreement with the GBM results, the data points should distribute around the line of $y = x$. We establish a criteria that the calculated results are considered acceptable if the data fall into the region between the line of $y = 0.75x$ and $y = 1.25x$ within 2~$\sigma$ errors.

\subsection{Comparison of $T_{90}$}

    \begin{figure}[!ht]
    \centering
    \subfloat[]{%
    \includegraphics[width=.45\textwidth]{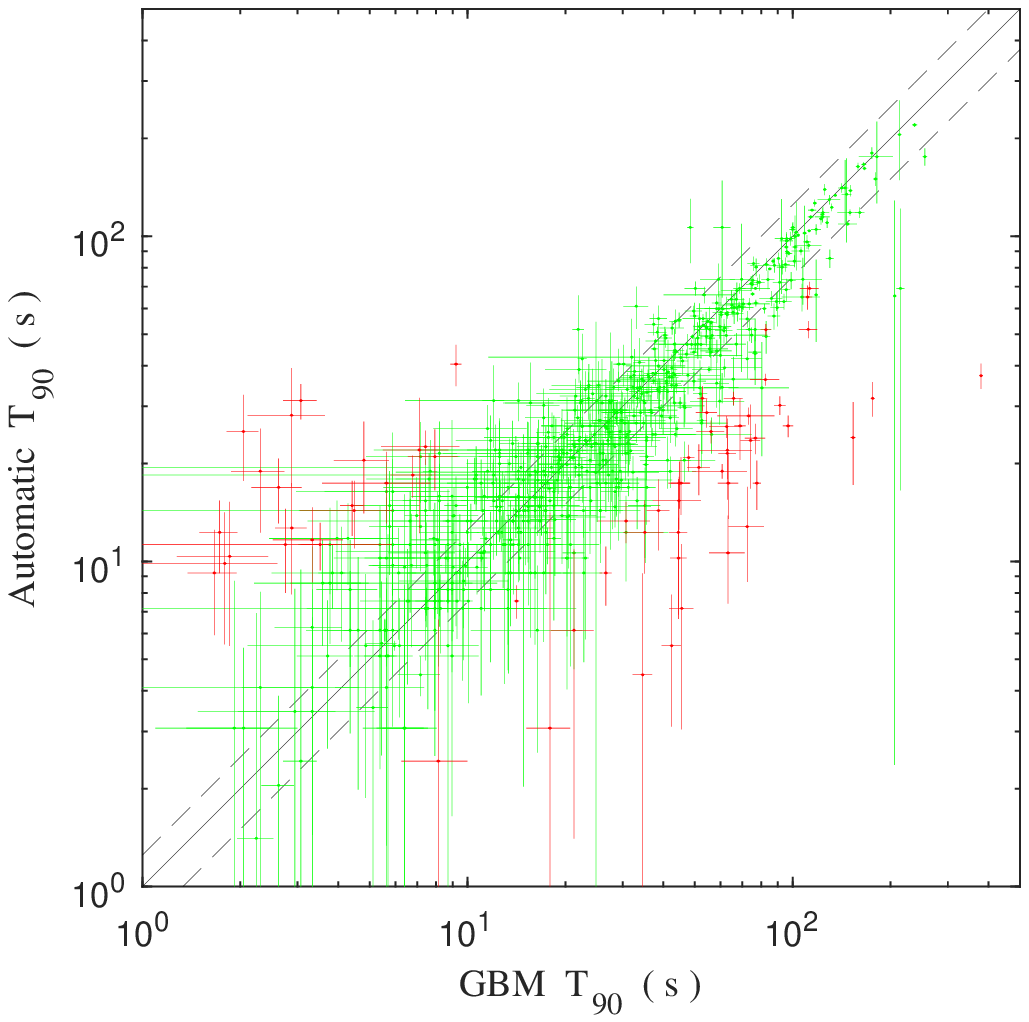}}\hfill
    \subfloat[]{%
    \includegraphics[width=.45\textwidth]{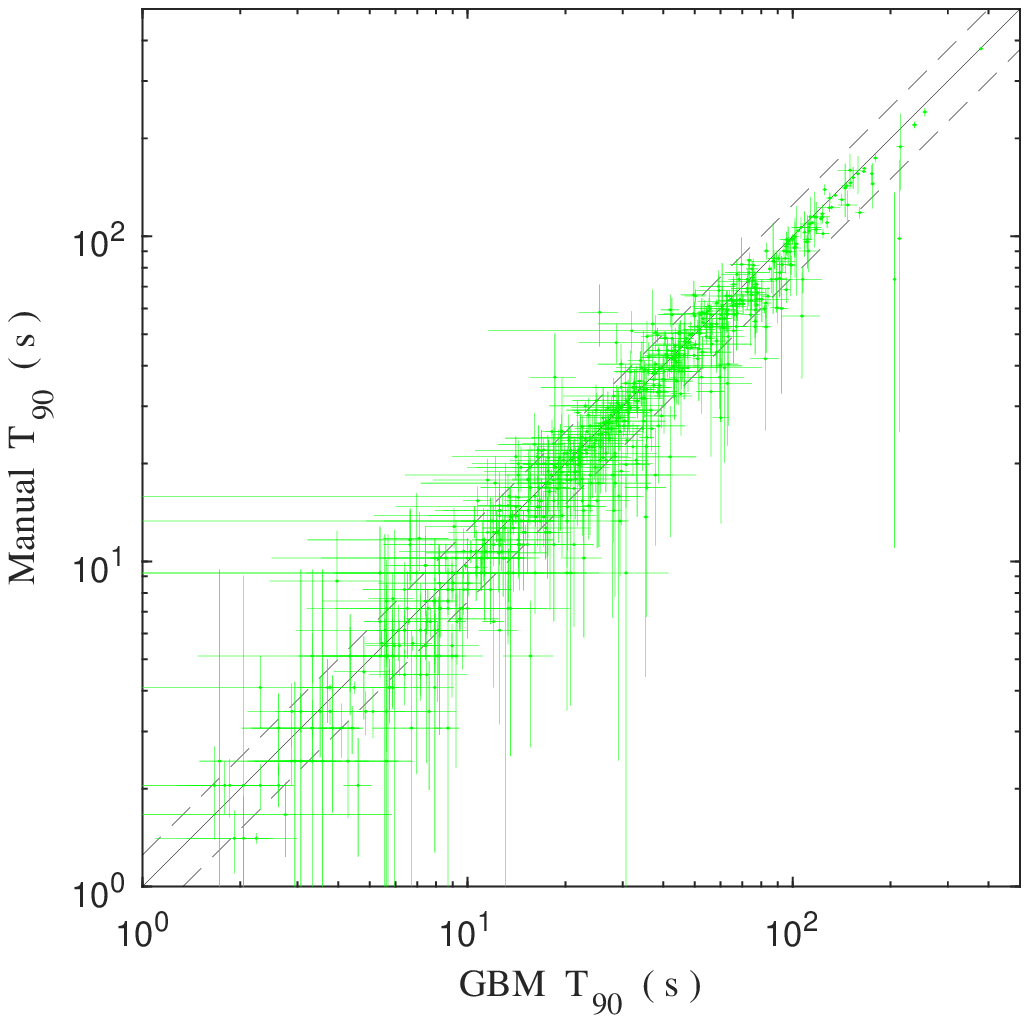}}\hfill
    \caption{ Comparison between $T_{90}$ obtained by our pipeline and given by GBM. Panel (a): automatically calculated $T_{90}$ against those given by GBM. Panel (b): manually calculated $T_{90}$ against those given by GBM. Green dots are GRBs compatible with GBM results while red dots are GRBs incompatible with GBM results according to our criterion (see the text). Diagonal solid lines are $y = x$, and the dashed lines are $y = 0.75x$ and $y = 1.25x$, respectively. }
    \label{raa20170280r1_fig5}
    \end{figure}

    According to this criterion, for $\sim90\%$ GRBs, our $T_{90}$~s are consistent with those of GBM (see Figure~\ref{raa20170280r1_fig5} (a)). After manually selecting the background and burst intervals, all GRBs are consistent with the GBM results, as shown in Figure~\ref{raa20170280r1_fig5} (b).

\begin{figure}[!ht]
    \centering
    \subfloat[]{%
    \includegraphics[width=.45\textwidth]{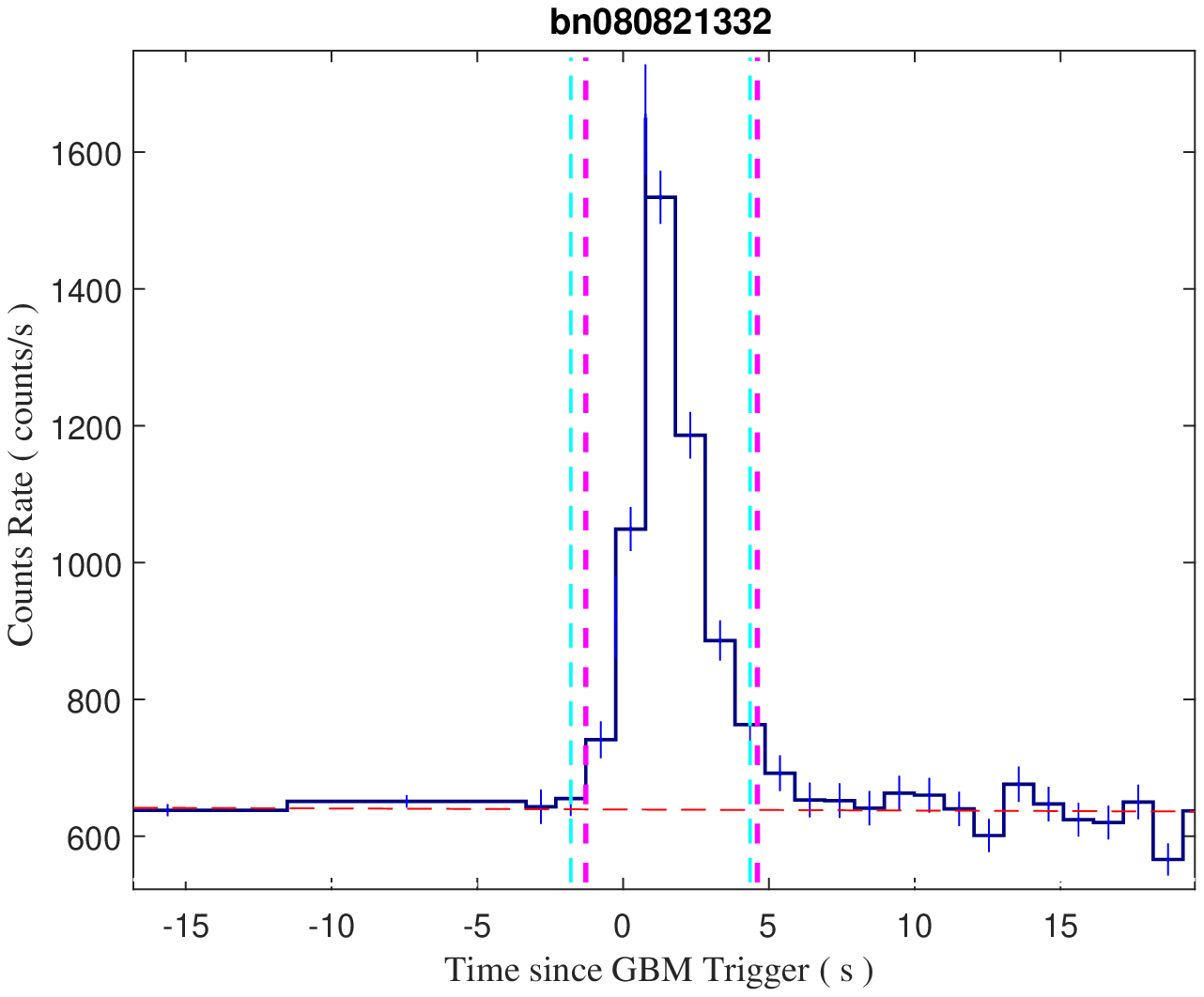}}\hfill
    \subfloat[]{%
    \includegraphics[width=.45\textwidth]{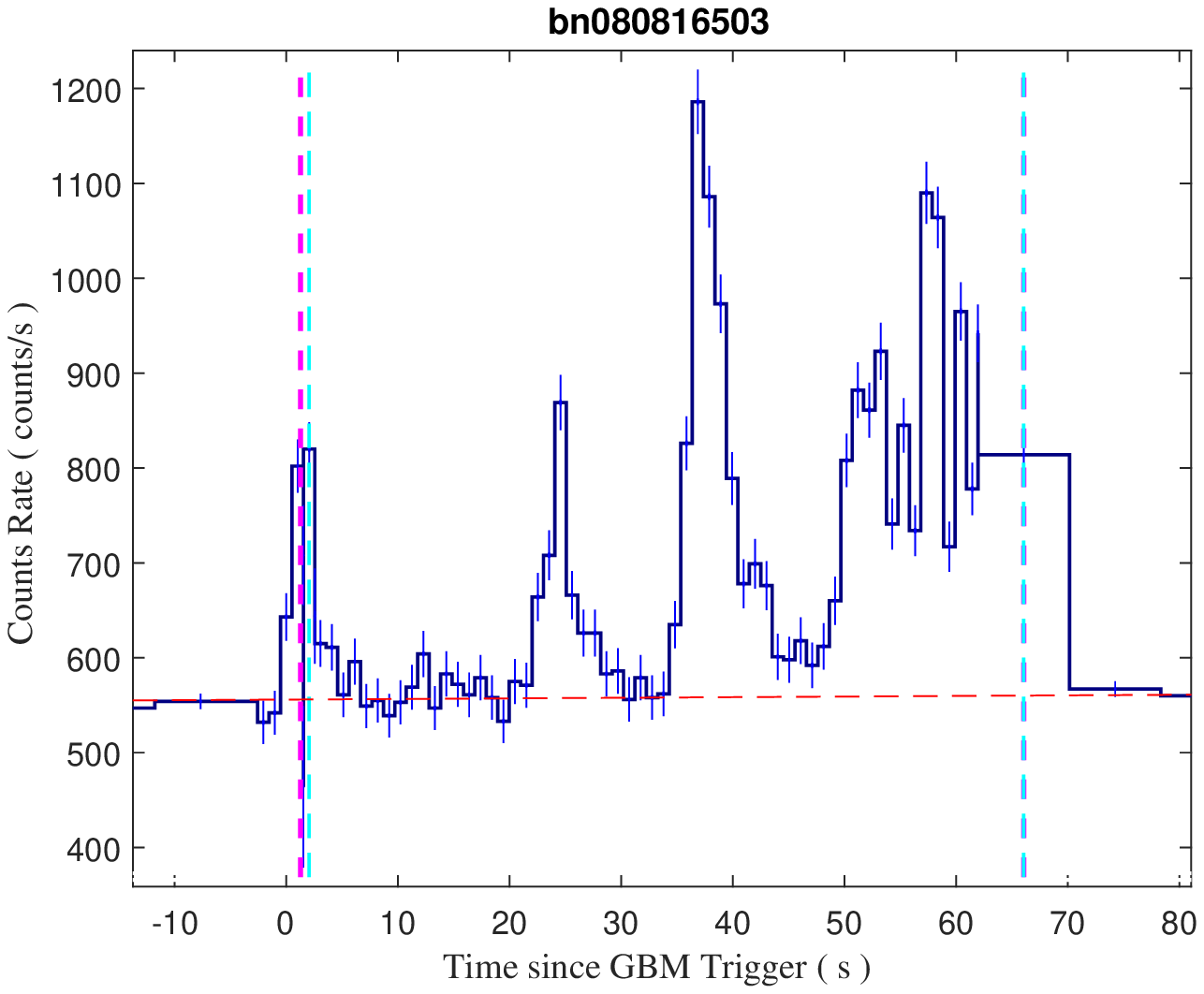}}\hfill
    \subfloat[]{%
    \includegraphics[width=.45\textwidth]{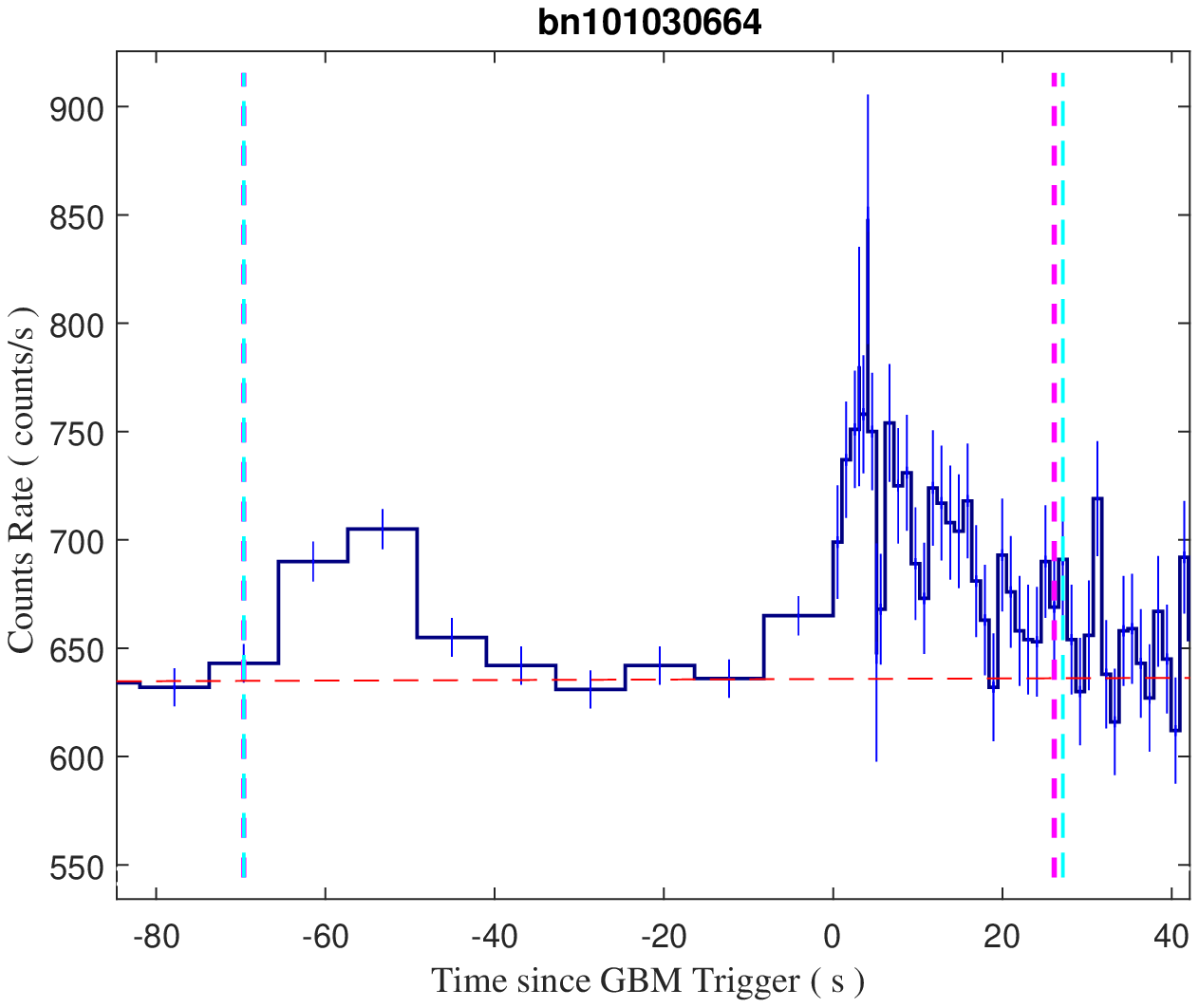}}\hfill
    \subfloat[]{%
    \includegraphics[width=.45\textwidth]{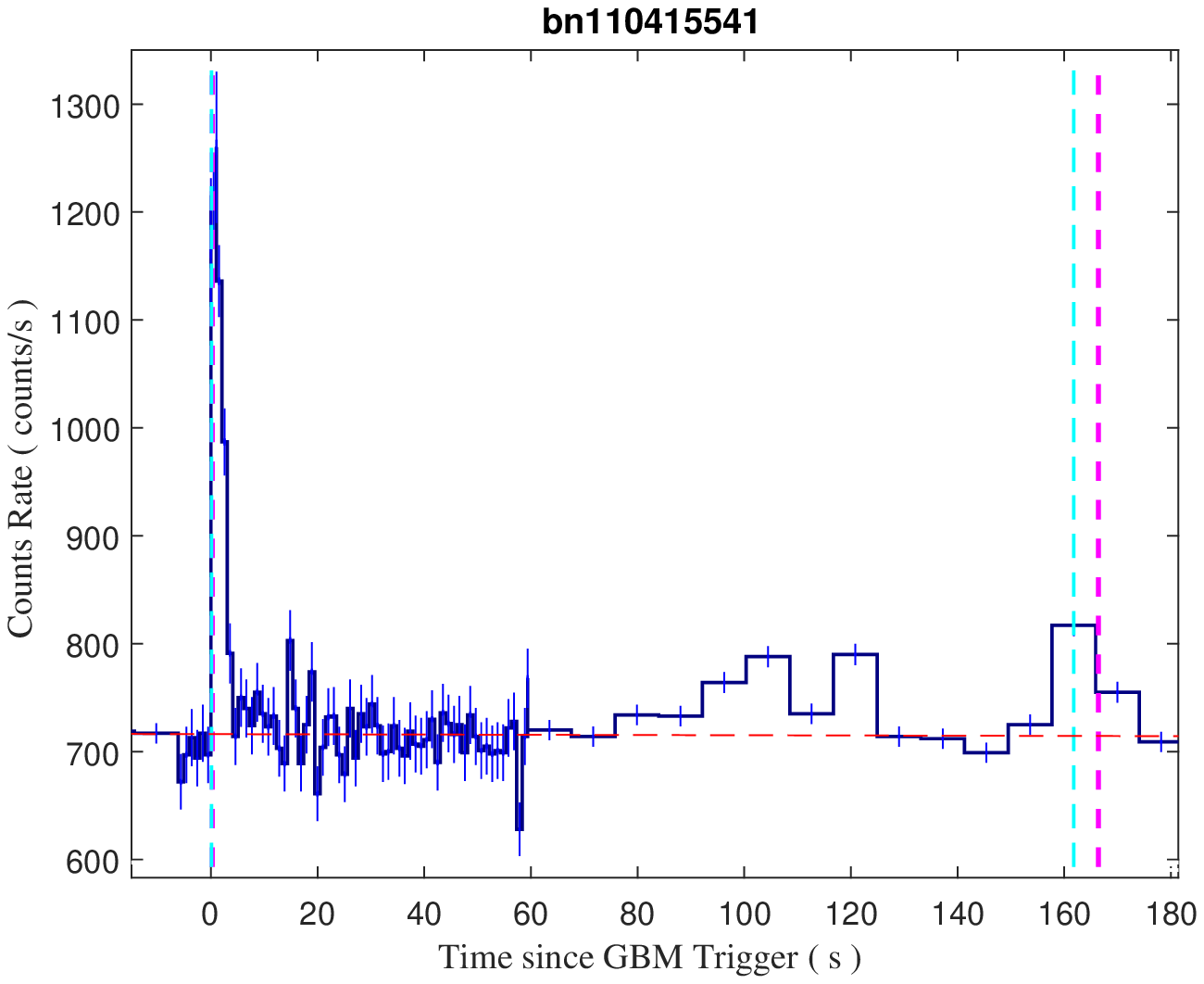}}\hfill
    \caption{ The typical cases with reasonable $T_{90}$ given by our automatic pipeline. The red dashed lines are the background fitting curves. The light blue dashed lines represent $T_{90}$ obtained automatically , while the magenta dashed lines represent the $T_{90}$ given by GBM. }
    \label{raa20170280r1_fig6}
    \end{figure}

\begin{figure}[!ht]
    \centering
    \subfloat[]{%
    \includegraphics[width=.45\textwidth]{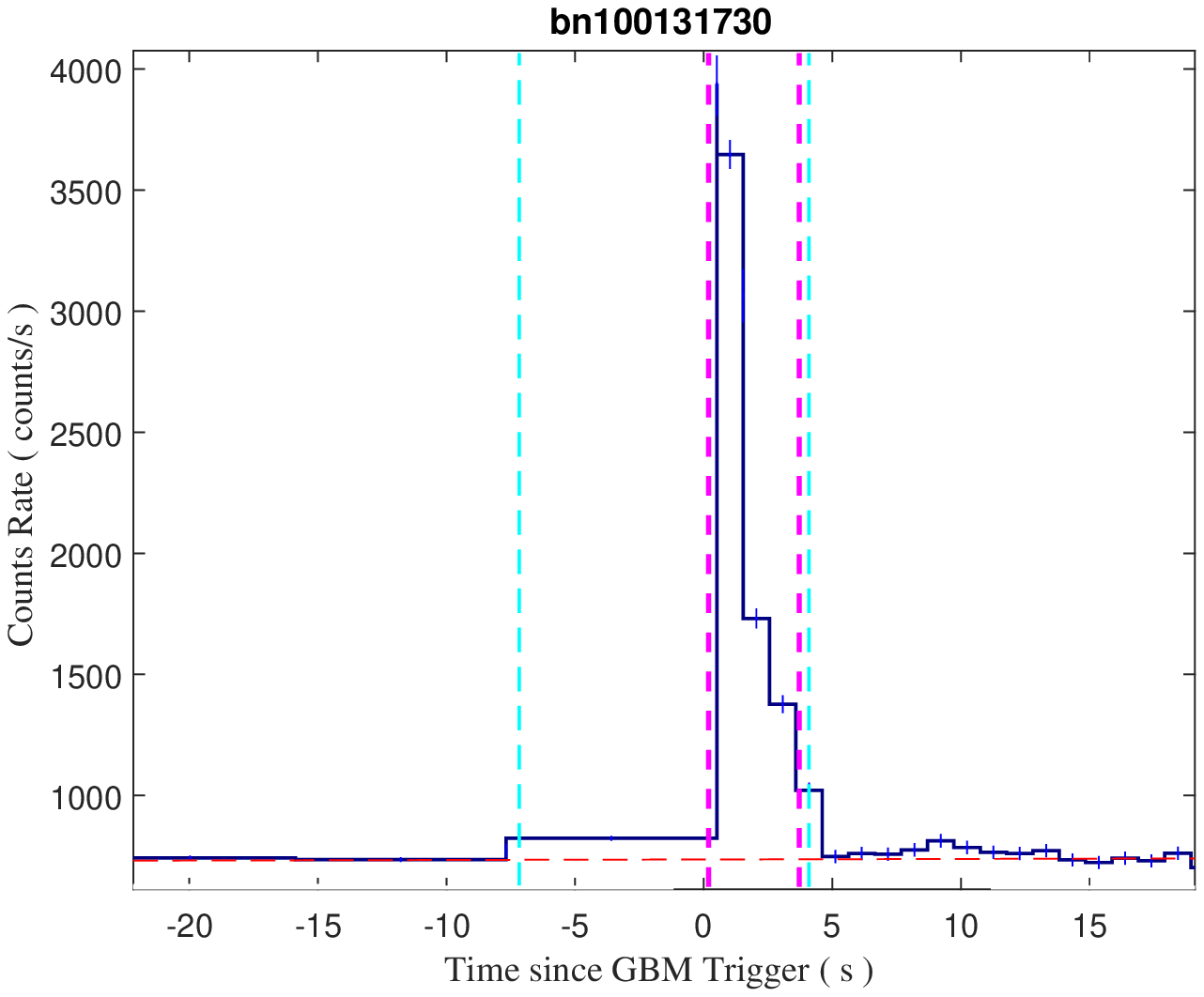}}\hfill
    \subfloat[]{%
    \includegraphics[width=.45\textwidth]{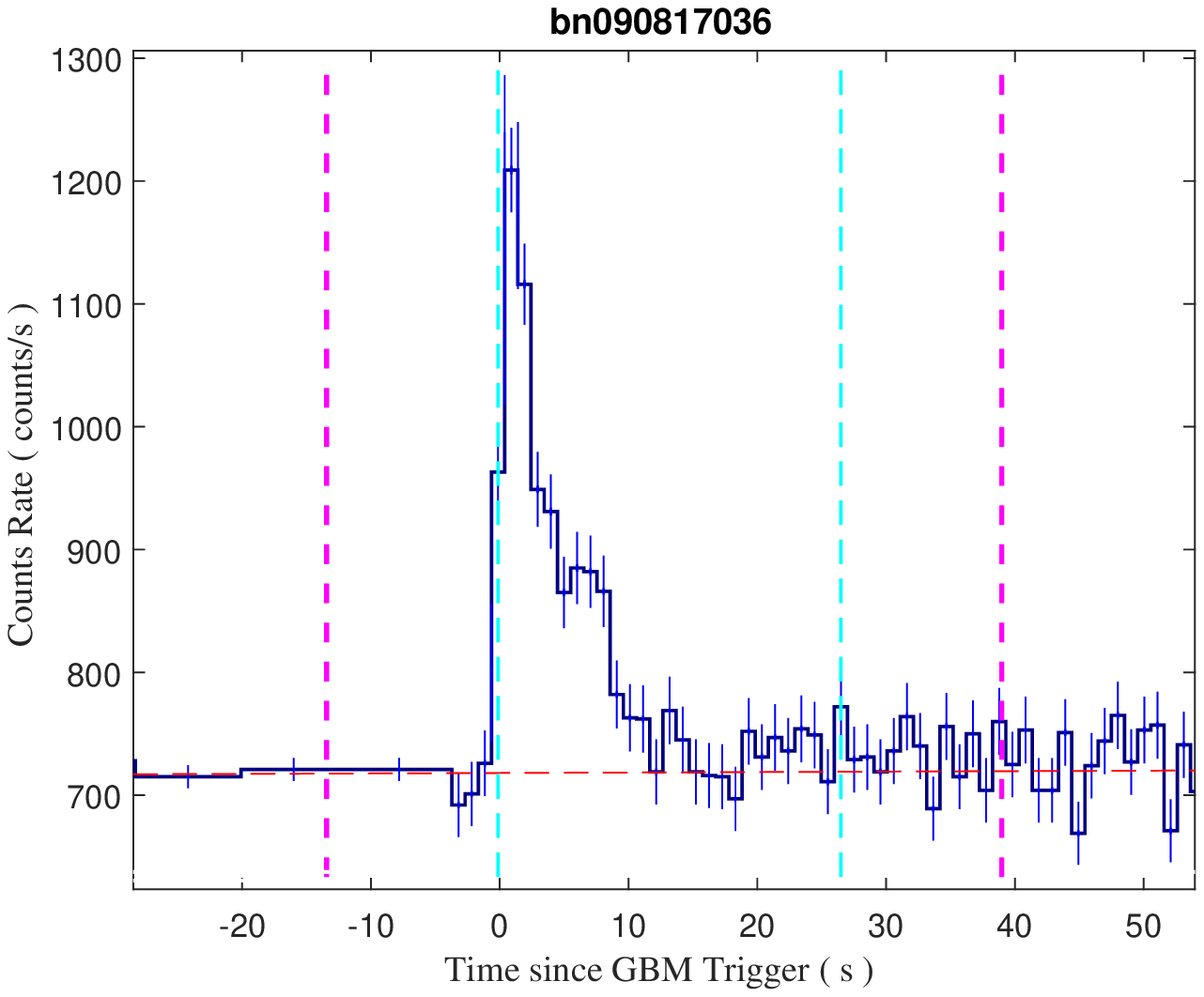}}\hfill
    \subfloat[]{%
    \includegraphics[width=.45\textwidth]{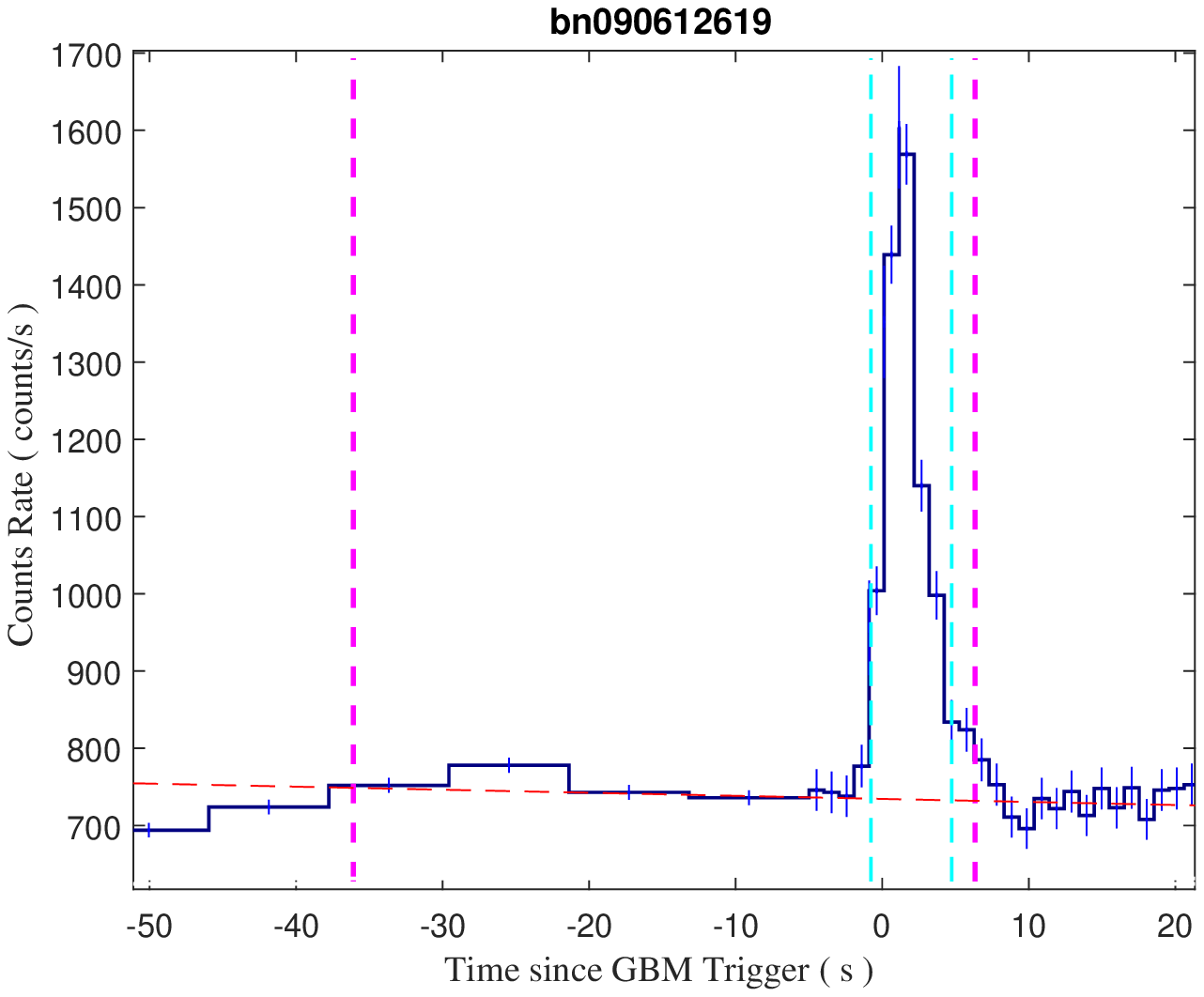}}\hfill
    \subfloat[]{%
    \includegraphics[width=.45\textwidth]{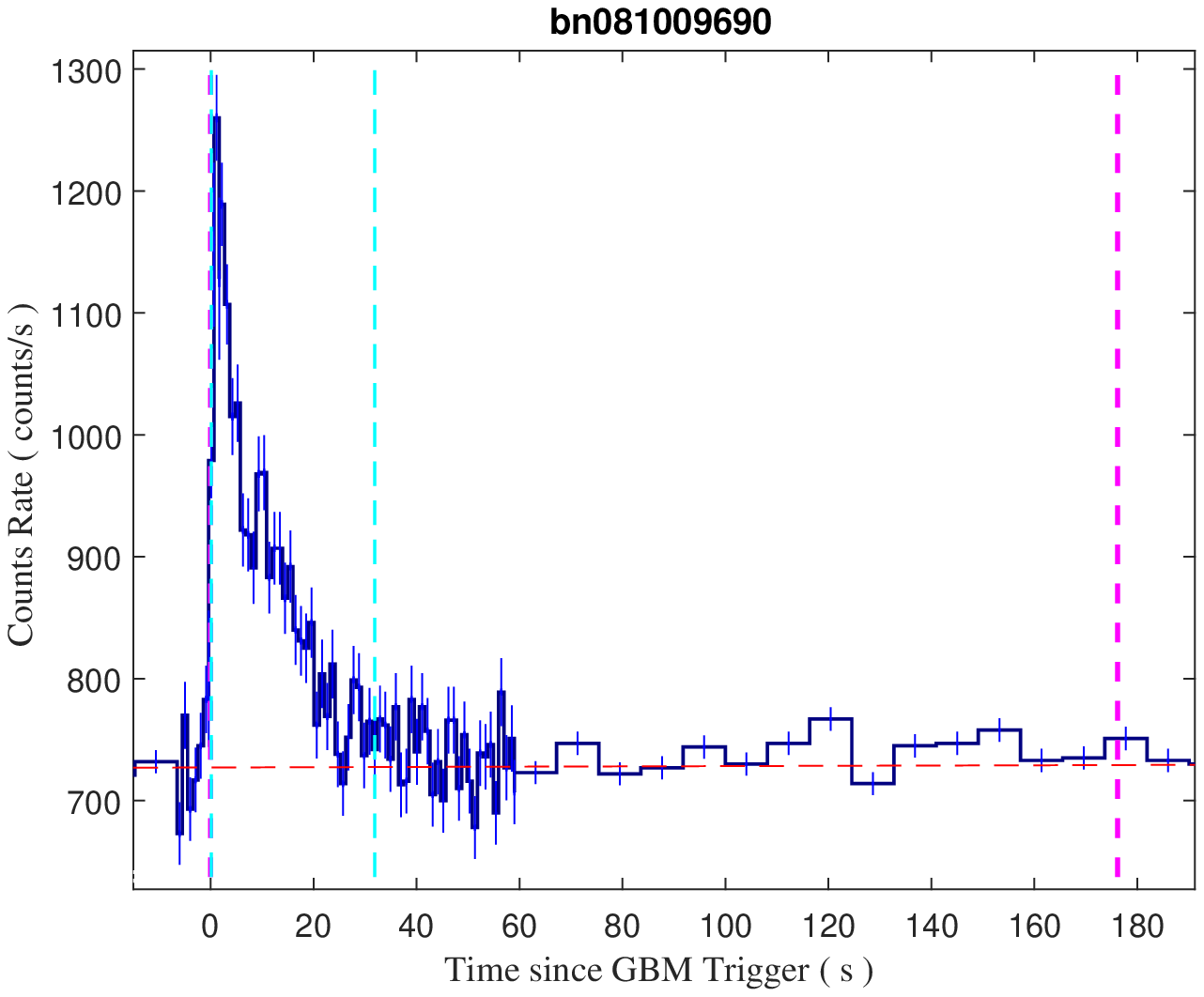}}\hfill
    \caption{ The typical cases with bad calculated $T_{90}$ given by our automatic pipeline. The definition of magenta lines and light blue lines are the same as Figure~\ref{raa20170280r1_fig6}. }
    \label{raa20170280r1_fig7}
    \end{figure}

    For most GRBs, our pipeline can give reasonable results as shown in Figure~\ref{raa20170280r1_fig6}, demonstrating that our pipeline can provide very good results for typical GRB light curves. However, a small fraction of GRBs are not compatible with the GBM catalog results (see the red dots in Figure~\ref{raa20170280r1_fig5}). Figure~\ref{raa20170280r1_fig7} shows the inconsistent cases of calculated $T_{90}$ given by our automatic pipeline.

    For some inconsistent cases, such as bn100131730 and bn090817036, the results of $T_{90}$ are shown in Figure~\ref{raa20170280r1_fig7} (a) and (b), respectively. Judging from the light curves created by the TRIGDAT data, the emission of bn100131730 near $T_0-5$~s could be a part of the burst, which significantly exceeds the background; the counts rate of bn090817036 near $T_0-10$~s and $T_0+30$~s cannot be distinguished from the background. It is worth to notice that the TRIGDAT data using to calculated $T_{90}$ have coarse temporal and spectral resolution. The TTE data and the location of the bursts are required to accurately identify the burst intervals. However, to ensure its efficiency, our pipeline doesn't include this part of the analysis.

    For other inconsistent cases, our pipeline fails to give the correct $T_{90}$ in automatic run (although manual processing can yield the right $T_{90}$). The main reason is that these GRBs present weak precursors or tails before or after the main peaks. So far, our automatic $T_{90}$ calculation cannot recognize precursors or tails with low significance, and sometimes even take them as background intervals by mistake, which leads to much shorter $T_{90}$ compared to the GBM results. Some examples can be seen in Figure~\ref{raa20170280r1_fig7} (c) and (d).

    Besides, for GRBs with $T_{90}$ less than 1~s, neither manual nor automatic $T_{90}$ calculations can give exact results due to the limitation of the time resolution of the TRIGDAT data. However, our pipeline can automatically give the upper limit of $T_{90}$ to determine whether a GRB is short or long, which plays an important role in the EM counterpart follow-ups. Acturally, it is easy to identify the duration of a GRB manually in this case.

\subsection{Comparison of Fluence}

\begin{figure}[!ht]
    \centering
    \subfloat[]{%
    \includegraphics[width=.45\textwidth]{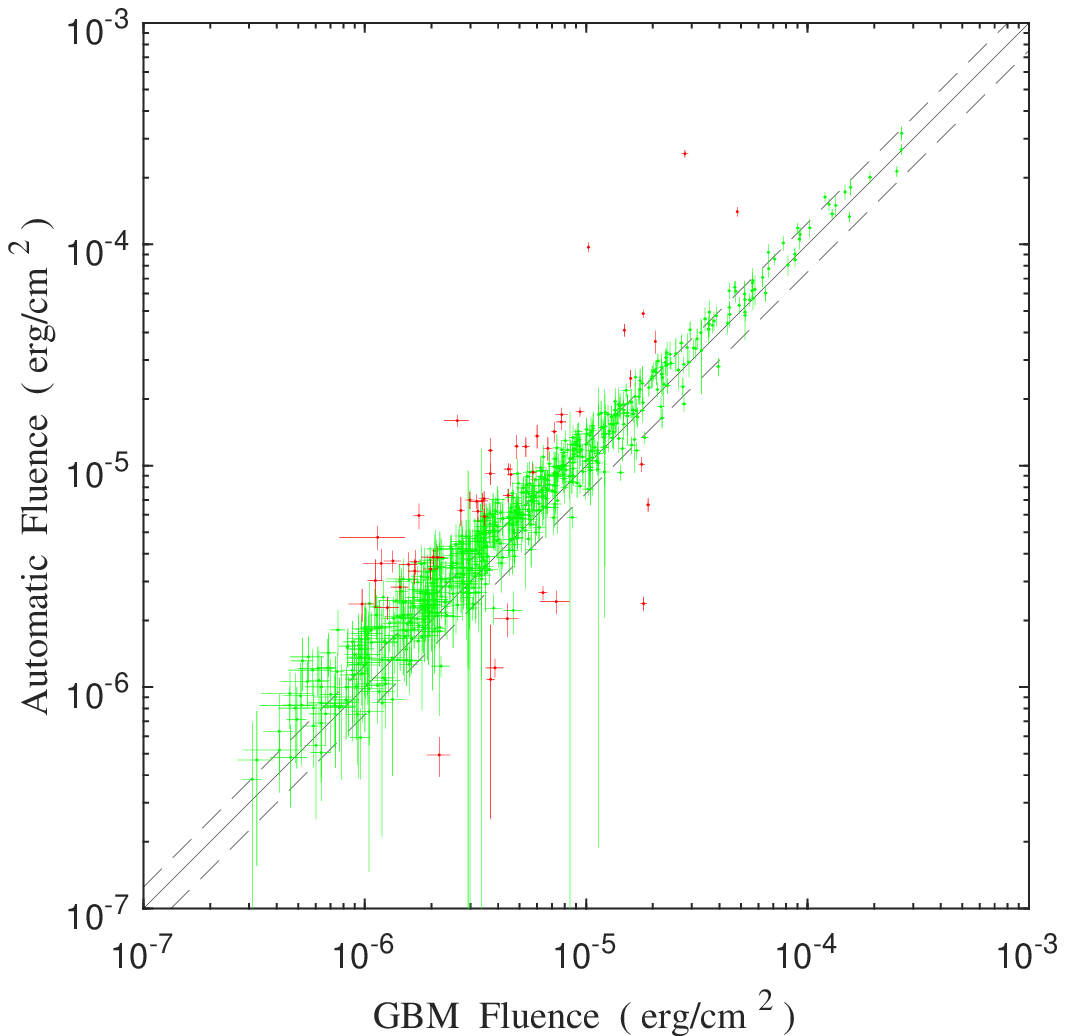}}\hfill
    \subfloat[]{%
    \includegraphics[width=.45\textwidth]{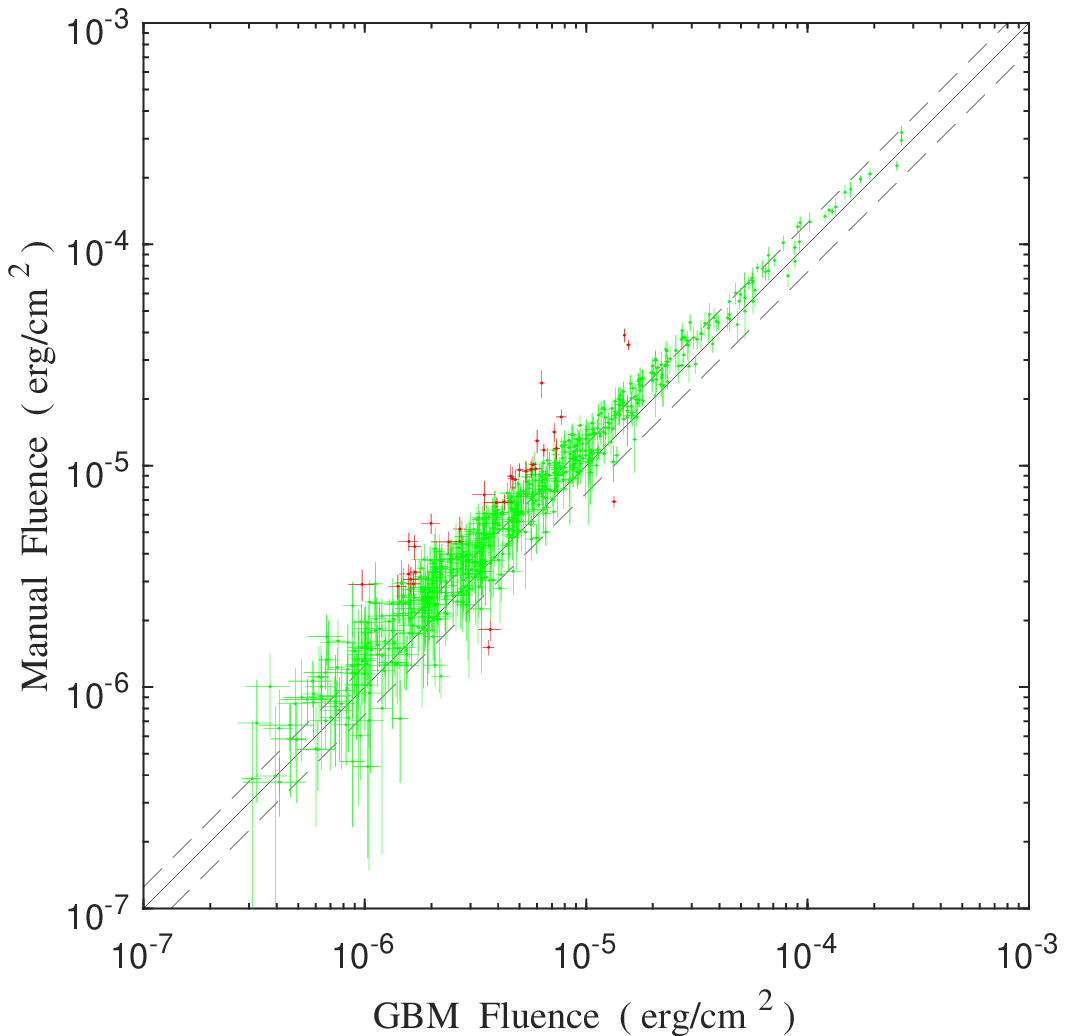}}\hfill
    \caption{ Comparison between this pipeline obtained fluence and those given by GBM. Panel (a): automatically calculated fluence against those given by GBM. Panel (b): manually calculated fluence against those given by GBM. The definition of green dots, red dots, solid lines and dashed lines are the same as Figure~\ref{raa20170280r1_fig5}. }
    \label{raa20170280r1_fig8}
    \end{figure}

    In order to compare the results of fluence with GBM, we used the same energy bands, time intervals and fitting model as those of the GBM catalog. As shown in Figure~\ref{raa20170280r1_fig8} (a), for 92\% GRBs, the fluence are consistent with the GBM results when the pipeline runs automatically. This indicates that the majority of fluence can be calculated reasonably well by our pipeline.

    It is worth mentioning that, for some GRBs, although the calculated $T_{90}$ of our pipeline is not very consistent with the GBM catalog, there is no significant difference on the main emission components between our calculation and GBM results, as shown in Figure~\ref{raa20170280r1_fig7}. Therefore, the fluence of these cases can also be calculated reliably, as tabulated in Table~\ref{Tab:T2}.

    \begin{table}[!ht]
    \begin{center}
    \caption[]{ The fluence results of 4 GRBs with bad $T_{90}$. }\label{Tab:T2}
    \begin{tabular}{ccccc}
        \hline\noalign{\smallskip}
                                                           &  bn100131730       &  bn090817036        &  bn090612619       & bn081009690        \\
        \hline\noalign{\smallskip}
        Our Automatic $T_{90}$ (s)                         & $ 11.26 \pm 1.95 $ & $ 26.62 \pm 6.82 $  & $ 5.50 \pm 2.24 $  & $ 31.74 \pm 3.89 $  \\
        GBM $T_{90}$ (s)                                   & $ 3.52 \pm 0.45 $  & $ 52.42 \pm 10.66 $ & $ 42.43 \pm 2.89 $ & $ 176.19 \pm 2.13 $ \\
        Fitting Models                                     &   CPL              &   CPL               &   CPL              & CPL     \\
        Our Automatic Fluence ($\times 10^{-6}$erg/cm$^2$) & $ 9.9 \pm 2.0 $    & $ 4.2 \pm 0.7 $     & $ 5.4 \pm 2.0 $    & $ 12.8 \pm 1.4 $    \\
        GBM Fluence ($\times 10^{-6}$erg/cm$^2$)           & $ 7.3 \pm 0.2 $    & $ 5.6 \pm 0.3 $     & $ 5.6 \pm 0.3 $    & $ 9.0 \pm 0.4 $     \\
    \noalign{\smallskip}\hline
    \end{tabular}
    \end{center}
    \end{table}

    The main factors for the unreasonable automatically calculated fluence are: (1) the low energy resolution of the TRIGDAT data; (2) the unreasonable selected background and burst intervals by our automatic pipeline; (3) the DRMs affected by the location accuracy and precision (Connaughton et al.~\citealp{conna2015}). After manually selecting the background intervals, the burst intervals and the fitting models, the coincidence rate increases to 95\%, as shown in Figure~\ref{raa20170280r1_fig8} (b). However, the spectral resolution for the TRIRDAT data is fixed --- only 8 energy channels compared with 128 channels in the CSPEC and TTE data. The accuracy and precision of the GBM location recorded in the TCAT data also have a weak effect on the fluence through influencing on the DRMs. Due to this intrinsic feature of the TRIGDAT data and some other problems such as the location-dependent DRMs, there are still $\sim5\%$ GRBs inconsistent with the GBM catalog spectral results.

\section{CONCLUSION AND DISCUSSION}

    In this paper, we developed an automatic low-latency pipeline for timing and spectral analysis based on {\it Fermi}/GBM near real-time data. Within $\sim20$~minutes after the GBM trigger, this pipeline could automatically accomplish timing and spectral analysis, and give some key parameters of GRBs, such as $T_{90}$ and fluence. In addition to the ability of completely automatic running, the pipeline allows for manually selecting background and burst intervals as well as spectral fitting models. For $\sim90\%$ GRBs, $T_{90}$ and fluence are consistent with the GBM catalog results within 2~$\sigma$ errors when the pipeline automatic runs. While for manually selecting the background intervals, the burst intervals and the fitting models, the coincidence rate can increase to $\sim95\%$ within 2~$\sigma$ errors.

    The main goal of this pipeline is to support the follow-up observation strategy of POLAR by providing the earliest $T_{90}$ and fluence after the GBM trigger. Low latency calculation of these properties could play an important role in joint and follow-up observations. If a GRB satisfy some criteria (e.g. long GRB with large fluence or short GRB), alerts for follow-up observations could be sent out at the earliest time. Meanwhile, the pipeline is also serving for the {\it Insight} Hard X-Ray Modulation Telescope ({\it Insight}-HXMT), the first Chinese X-ray space telescope which is able to detect GRBs and EM counterparts of GW events (Li et al.~\citealp{litip2018}). This pipeline has been used by POLAR and {\it Insight}-HXMT, and will be open to the astronomical community.

    The detection and research of GWs and their prompt EM counterparts have increased significantly of late. Previous studies suggest that the short-duration gamma-ray bursts are likely to be generated by the mergers of binary neutron stars (BNSs) or a neutron star and a black hole (Fern\'{a}ndez \& Metzger~\citealp{ferna2016}). Thus they are usually considered to be the high-energy EM counterparts of GWs. Recently, LIGO/Virgo, {\it Fermi}/GBM and INTEGRAL/SPI-ACS have jointly found the first GW event GW170817 and its prompt EM counterpart GRB 170817A (Abbott et al.~\citealp{abbot2017a}, ~\citealp{abbot2017b}, ~\citealp{abbot2017c}) originated from the first BNS merger.

    After receiving the GBM trigger notice, this pipeline immediately processed the near real-time data of GRB 170817A, and provided the preliminary information in 21 minutes after the GBM trigger. It reported that the fluence of the GRB 170817A is $(3.3 \pm 0.7)\times 10^{-7}$ erg/cm$^2$, which is consistent with $(2.8 \pm 0.2)\times 10^{-7}$ erg/cm$^2$ (Goldstein et al.~\citealp{golds2017}) at 1~$\sigma$ level.

    It is anticipated that more GW events generated by mergers of BNSs and their EM counterparts will be found in the near future. With our pipeline, primary timing and spectral information about GRBs can be calculated and reported rapidly, which is of great importance for the follow-up EM counterpart observations.

    The main advantage of this pipeline is low-latency and automatic. For most of GRBs, our pipeline can give quite reliable results, though it does not work well for a small fraction of bursts with very special light curves. We emphasize, however, this may be partially, if not totally, due to the intrinsic low temporal and spectral resolution of the near real-time data (TRIGDAT, TCAT) that the pipeline used.

    Continuous improvements of this pipeline are undergoing. The first step is to estimate the in-flight background of GBM using a direction dependent background fitting (DDBF) method (Sz\'{e}csi et al.~\citealp{szecs2013}) or to estimate the background level from the last 30 or 60 orbits (Fitzpatrick et al.~\citealp{fitzp2011}). The second step is to introduce the deep learning method (Lecun et al.~\citealp{lecun2015}) to intelligently distinguish the background and burst intervals and to select the suitable fitting models. With these improvements, the background and burst intervals of GRBs could be distinguished better by the pipeline and more reliable results of temporal and spectral analysis can be obtained.

\normalem
\begin{acknowledgements}

    We thank the anonymous reviewer for helpful comments and suggestions. This work is supported by the Strategic Priority Research Program of the Chinese Academy of Sciences (Grant No. XDB23040400), the National Natural Science Foundation (Grant No. 11403026, 11503028 and 11673023) and the Basic Research Program (973 program) of China under (Grant No. 2014CB845800). BBZ acknowledge the support from the National Thousand Young Talents program of China. We gratefully acknowledge the support of the collaboration team of POLAR, a project funded by China National Space Administration (CNSA), the Chinese Academy of Sciences (CAS) and the University of Geneva (UNIGE). The authors also would like to thank Jing Jin, Jinlu Qu, Xiaobo Li, Youli Tuo, Zijian Li, Mingyu Ge, Jinyuan Liao, Guangcheng Xiao, Yue Huang and Chengkui Li for their suggestions on data analysis and the revision of this paper.

\end{acknowledgements}

\end{document}